\documentclass[%
 reprint,
 amsmath,amssymb,
 aps,
]{revtex4-1}

\usepackage{graphicx}
\usepackage{dcolumn}
\usepackage{bm}

\begin{document}

\preprint{APS/123-QED}

\title{Gravitational lensing in plasma: \\
Relativistic images at homogeneous plasma}

\author{Oleg Yu. Tsupko}
 \email{tsupko@iki.rssi.ru}
\author{Gennady S. Bisnovatyi-Kogan}%
 \email{gkogan@iki.rssi.ru}
\affiliation{%
 Space Research Institute of Russian Academy of Sciences, Profsoyuznaya 84/32, Moscow 117997, Russia\\
 National Research Nuclear University MEPhI, Kashirskoe Shosse 31, Moscow 115409, Russia\\
}%

\date{\today}

\begin{abstract}
We investigate the influence of plasma presence on relativistic images formed by Schwarzschild black hole lensing. When a gravitating body is surrounded by a plasma, the lensing angle depends on a frequency of the electromagnetic wave due
to refraction properties, and the dispersion properties of the light propagation in gravitational field in plasma. The last effect leads to difference, even in uniform plasma, of gravitational deflection angle in plasma from vacuum case. This angle depends on the photon frequency, what resembles the properties of the refractive prism spectrometer. Here we consider the case of a strong deflection angle for the light, traveling near the Schwarzschild black hole, surrounded by a uniform plasma. Asymptotic formulae are obtained for the case of a very large deflection angle, exceeding $2\pi$. We apply these formulae for calculation of position and magnification of  relativistic images in a homogeneous plasma, which are formed by the photons performing one or several revolutions around the central object. We conclude that the presence of the uniform plasma increases the angular size of relativistic rings or the angular separation of point images from the gravitating center. The presence of the uniform plasma increases also a magnification of relativistic images. The angular separation and the magnification become significantly larger than in the vacuum case, when the photon frequency goes to a plasma frequency.

\begin{description}
\item[PACS numbers] 04.20.-q - 95.30.Sf - 98.62.Sb - 94.20.ws


\end{description}
\pacs{04.20.-q - 95.30.Sf - 98.62.Sb - 94.20.ws}
\end{abstract}

\pacs{04.20.-q - 95.30.Sf - 98.62.Sb - 94.20.ws}
\maketitle


\section{Introduction}

A theory of gravitational lensing is well developed for the light propagation in vacuum, see e.g.\cite{GL,GL2}. The theory usually deals with a
geometrical optics in vacuum and uses a notion of the deflection angle. Basic assumption used in a gravitational lensing theory is
an approximation of a weak deflection angle of a photon. The deflection is small if the impact parameters for the incident photon $b$ is
much greater than the Schwarzschild radius $R_S$ of the gravitating mass (lens). In this approximation the trajectory of the photon is
almost straight line with a weak deflection by the Einstein angle $\hat{\alpha} = \frac{2 R_S}{b} = \frac{4M}{b}$, $G=c=1$. For most  astrophysical situations related to the
gravitational lensing, the above weak deflection condition is well satisfied, and all observational data are related to this case.
The theory of a weak gravitational lensing is  developed in \cite{GL,GL2,Perlick2004b,Perlick2010}. The deflected light rays change an apparent position of the
source, its shape, may form several images of the source (see Figure 1).\\

One way to expand the consideration is to go beyond the weak deflection limit. If the photon impact parameter is close to its
critical value, a photon which goes from infinity can perform several turns around the central object and then go to infinity. In
this case deflection angle is not small.
An exact expression for the deflection angle at arbitrary impact parameter was obtained by Darwin \cite{Darwin1959}, who managed to express the result via elliptic integrals.
Photons from a distant source which undergo one or several loops around the central object (lens), and then go to observer form
images, which are called relativistic images \cite{Virbhadra2000} (see Figure 2). When the lens, source and the observer are  situated on one line, these images form concentring rings around the lens. In the opposite case the relativistic images could be visible as two infinite sequences of "spots", situated on one line at both sides of the lens, and converging to the radius which value is of the order of the Schwarzschild radius of the lens.
Using an exact expression for the deflection angle, Virbhadra and Ellis \cite{Virbhadra2000}
calculated numerically the positions and the magnifications of the relativistic images for the Schwarzschild space-time.
Frittelli, Kling, Newman \cite{Frittelli2000} considered the exact lens equation for the Schwarzschild metric, and obtained
solutions in the form of integral expressions. Exact gravitational lens equation in the spherically symmetric and static space-time
was also considered by Perlick \cite{Perlick2004a}.

In the case of a very large deflection, the value of the bending angle in the Schwarzschild
metric can be written in a simple analytical form. In a strong deflection approximation, the deflection angle diverges
logarithmically while the impact parameter approaches its critical value. The analytic  expression was first derived by
Darwin \cite{Darwin1959}, and is usually referred as a strong deflection limit. Bozza et al \cite{Bozza2001} applied this analytical expression for the investigation of the relativistic images, finding the positions and the magnifications of the relativistic images in the Schwarzschild metric. It was shown in \cite{Bozza2002} that for a general spherically symmetric space-time the deflection angle diverges logarithmically when the minimum impact parameter is reached. Relativistic rings for a Schwarzschild black hole lens were considered in more details in \cite{BKTs2008}. Strong deflection approximation was argued in \cite{Virbhadra2009}.

Relativistic images in vacuum in different types of metric, and alternative theories of gravitation have been extensively studied in \cite{IyerPetters, Eiroa2002, Keeton, Cite1,Cite2,Cite3,Virbhadra2009, PerlickMG, Perlick2002, Perlick2004b,Perlick2010, EiroaSendra}. Existence of photon spheres plays important role in gravitation lensing \cite{Virbhadra2000, Virbhadra2001, Virbhadra2002}. In any space–time containing a photon sphere, gravitational lensing will give rise to relativistic images. \cite{Virbhadra2000, Virbhadra2001, Virbhadra2002}. The role of a scalar field on image positions and magnifications was first shown in \cite{Virbhadra1998} and \cite{Virbhadra2002}. In \cite{Virbhadra1998} a static and circularly symmetric lens characterized by the mass and scalar charge parameters is constructed. In \cite{Virbhadra2002} authors model massive dark objects in galactic nuclei as spherically symmetric static naked singularities in the Einstein massless scalar field theory, and study the resulting gravitational lensing in detail. Time delay is also important lensing feature, see for example the paper \cite{Virbhadra2008} where the time delay expression for a general static spherically symmetric metric is analysed.\\

Another way to expand the usual gravitational lens theory is to consider a medium instead of a vacuum. In a space the light rays
propagate through the plasma, so the main interest is to consider how the deflection angle changes in presence of the plasma.

This problem is not new. The consideration of the photon deflection by a gravitating center, in presence of plasma around it, was
considered in classical book of problems \cite{zadachnik}, for more details see our previous works \cite{BKTs2009},
\cite{BKTs2010}. In application to gravitational lensing it was discussed in \cite{BliokhMinakov} where gravitational
lensing by the gravitating body with surrounding spherically symmetric plasma distribution was considered.

Working in the frame of a geometrical optics, what is usual for gravitational lensing, we can characterize the properties of
the medium by its refractive index. In an inhomogeneous medium the refractive index depends explicitly on the space
coordinates, and the photon moves along a curved trajectory, what is called as a refraction. The effect of refractive deflection of the
light has no relation to relativity and gravity, and takes place due to non-homogeneity of the media. Bliokh and Minakov \cite{BliokhMinakov} have
performed study of gravitational lensing in plasma in the approximation,
when a deflection angle is
just a sum of two separate effects: a vacuum
deflection due to gravitation of a point mass, and a refractive deflection due to the non-homogeneity of the plasma.

A self-consistent approach to the light propagation in the gravitational field, in presence of a medium, was developed in a classical book of Synge \cite{Synge}.
For discussion and application of the Synge's general relativistic Hamiltonian theory for the geometrical optics see \cite{Bicak,
Krikorian1985,Krikorian1999}. Similar approach for the propagation of the light rays, in presence of both gravity and
plasma, was developed by Perlick \cite{Perlick2000}, where general formulae for the light deflection angle in the Schwarzschild and
Kerr metrics, in presence of plasma, have been obtained in the integral form.

In \cite{BKTs2009}, \cite{BKTs2010} we have found, that in absence of refraction, in a uniform {\it dispersive} medium, when the refractive index depends on the photon frequency, the gravitational deflection is qualitatively different from
the vacuum case, and the deflection angle depends on the photon frequency (see Figure 3).
We used a general theory, developed in \cite{Synge} for the geometrical optic in the curved space-time, in a dispersive medium, where the refractive index is considered as a scalar, depending only on the photon frequency, and on the local parameters of the matter.
Applying the general theory to  plasma, that is a dispersive medium, and its refractive index depends on the photon frequency,
we have obtained  a simple analytical formula for the light deflection in a Schwarzschild metric, in presence of a homogeneous
plasma  \cite{BKTs2009}. In \cite{BKTs2010} we developed a more general approach, considering an inhomogeneous plasma, where a nongravitating refractive deflection was also taken into account. All the results have been obtained in the approximation of a small deflection angle.\\

In this work we develop
a model of gravitational lensing in presence of plasma without the restriction of a small deflection angle, and obtain analytical formulae
in the strong deflection angle limit. The paper is organized as follows. In
Section 2 we present a short review of the theory of a gravitational lensing in vacuum and plasma, what is necessary to make text more clear. In
Section 3 we derive the exact expression for the deflection angle of the photon moving in the Schwarzschild metric and in the
plasma with a spherically symmetric distribution of density. In Section 4 and 5 we consider the case of homogeneous plasma and express
the formula for deflection via elliptic integrals. In Section 6 we discuss a possibility of the finite orbits of the photons in a homogeneous plasma. In Section 7 we find a critical value for the closest approach distance at which a photon from the infinity remains on the circular orbit with an infinite number of circles around the center.
This critical value depends on the ratio of a photon and plasma
frequencies. The deflection angle goes to infinity when the distance of closest approach goes to this value. In Section 8 we
derive an analytical formula for the photon deflection angle in the Schwarzschild metric, in a homogeneous plasma, in the limit of a strong
deflection. In Section 9 and 10 we calculate the angular positions and magnification factors of the relativistic images in the case of lensing in a homogeneous plasma. In Section 11 we conclude and discuss our results.

\section{Gravitational deflection of photons in vacuum, and in plasma}

The system of units used below is

\begin{equation}
G=c=1, \quad \mbox{the Schwarzschild radius} \;\; R_S=2M.
\end{equation} \\

\textit{Gravitational deflection in vacuum, weak deflection limit}
The photon deflection angle in vacuum, in the
Schwarzschild metric with a given mass M, is determined, for small deflection
angles $\hat{\alpha} \ll 1$, by a formula
\begin{equation} \label{einstein}
\hat{\alpha} = \frac{2 R_S}{b} = \frac{4M}{b} ,
\end{equation}
where $b$ is the impact parameter, and $b \gg R_S$, $R_S = 2M$ is the Schwarzschild radius (see Figure 1). This angle does
not depend on the photon frequency. \\

\textit{Gravitational deflection in vacuum, exact expression.}
Exact expression for the deflection angle $\hat{\alpha}$ in the case of a motion in the Schwarzschild metric can be derived from equations
determining the photon orbit, see, for example, \cite{MTW, BKTs2008}. For a given mass $M$, the deflection angle of a photon
is a function of the radius of the closest approach $R$, and is represented by the integral

\begin{equation}
\label{vacuum-exact} \hat{\alpha} = 2 \int \limits_R^\infty \frac{dr}{r^2 \sqrt{\frac{1}{b^2} - \frac{1}{r^2} \left( 1- \frac{2M}{r}
\right) }} - \pi \, .
\end{equation}
The  impact parameter $b$, corresponding to  the distance of the closest approach $R$ is written as

\begin{equation}
\label{b&R} b^2 = \frac{R^3}{R-2M}.
\end{equation}
 The expression of the integral (\ref{vacuum-exact}) in terms of elliptical integrals was first obtained in \cite{Darwin1959}.
The expansions of exact deflection angle in powers of $M/R$ and $M/b$ is given in \cite{Keeton}. In the case of a large
impact parameters $b \gg 3\sqrt{3} M$ (weak deflection limit) we can neglect a difference between the impact parameter
and the distance of closest approach, and write formula (\ref{einstein}) either with $b$ or $R$. A form of the
formula (\ref{einstein}) in textbooks depends usually on a way of derivation (compare, for example, \cite{LL2} and \cite{MTW}).\\

\textit{Gravitational deflection in vacuum, strong deflection limit.} The critical value $b_{cr} = 3\sqrt{3} M$ corresponds to
the distance of closest approach equal to $R = 3 M$. If the value of the impact parameter is close to the critical value $0 < b/M
- 3\sqrt{3} \ll 1$, then the photon makes one or several turns around the black hole near the radius $r = 3M$, and the number of turns goes to infinity at $b\rightarrow b_{cr}$ (see Figure 2). In the case of a strong deflection limit the deflection angle can be written in the form
\cite{Darwin1959}, \cite{Bozza2001}
\begin{equation}
\label{vacuum-alpha-R}  \hat{\alpha} = - 2 \ln \frac{R-3M}{36(2-\sqrt{3}) M} - \pi \, ,
\end{equation}
or, as a function of the impact parameter $b$, in the form \cite{Bozza2001}, \cite{BKTs2008}
\begin{equation}
\label{vacuum-alpha-b}  \hat{\alpha} = - \ln \left( \frac{b}{b_{cr}} - 1 \right) + \ln[216(7-4\sqrt{3})] - \pi \, .
\end{equation}\\

\textit{Deflection of light rays in presence of a gravity and plasma, weak deflection limit.} Let us consider a plasma in a static weak gravitational field,  with a refractive index
\begin{equation} \label{plasma-n}
n^2 = 1 - \frac{\omega_e^2}{[\omega(r)]^2} \, , \quad \omega_e^2 = \frac{4 \pi e^2 N(r)}{m} \, .
\end{equation}
Here $\omega(r)$ is the frequency of the photon, which depends on the space coordinate $r$ due to the presence of a
gravitational field (gravitational red shift). We denote $\omega(\infty) \equiv \omega$, $e$ is the charge of the electron, $m$
is the electron mass, $\omega_e$ is the electron plasma frequency, $N(r)$ is the electron concentration in an inhomogeneous plasma.

We have shown for the first time \cite{BKTs2009, BKTs2010} that due to dispersive properties of plasma even in the homogeneous plasma the gravitational
deflection differs from vacuum deflection angle, and gravitational deflection angle in plasma depends on frequency of the photon as

\begin{equation} \label{main-res}
\hat{\alpha} = \frac{R_S}{b} \left( 1 + \frac{1}{1 - (\omega_e^2
/ \omega^2)} \right).
\end{equation}
This formula is valid under the condition of smallness of $\hat{\alpha}$. The presence of plasma increases the gravitational
deflection angle. Formula is valid only for $\omega > \omega_e$, because the waves with $\omega< \omega_e$ do not propagate in
the plasma. Under $\omega_e=0$ (concentration $N(r)=0$) or $\omega \rightarrow \infty$ this formula turns into the deflection
angle for vacuum $2R_S/b$.

In a homogeneous plasma the photons of smaller frequency, or larger wavelength, are deflected by a larger angle by the gravitating
center. The effect of difference from the vacuum case in the gravitational deflection angles is significant for longer wavelengths,
when $\omega$ is approaching $\omega_e$. In the space it is possible only for the radio waves. Therefore, the gravitational
lens in plasma acts as a radiospectrometer \cite{BKTs2009}. This effect has a general relativistic nature, in combination with
the dispersive properties of plasma. We should also emphasize that the plasma is considered here as a medium with a given index
of refraction, and this formula does not take into account selfgravitation of a plasma particles.

The observational effect of the frequency dependence may be demonstrated on the example of the Schwarzschild point-mass lens.
Instead of two point-like images with complicated spectra, we will have two line images, formed by the photons with different
frequencies, which are deflected by different angles (see Figure 3). Regretfully, all plasma effects in the gravitational lensing are
very small and their observations are scarcely possible at the moment. Different kinds of absorption, as well as refraction
properties of the nonuniform plasma contaminate this effect. Special conditions should exist for possibility to detect it
observationally. The optical depth due to Thomson scattering and free-free absorption during the process of gravitational lensing in a plasma have been estimated in \cite{BKTs2010}.

In the paper \cite{BKTs2010} we have  considered  the gravitational lensing of radiowaves by the point or the spherical body in
presence of a homogeneous and  non-homogeneous plasma,
without selfgravitation, together with the refractive deflection due to the plasma inhomogeneity (non-relativistic effect), and comparison of these two effects have been there performed. All these results have been obtained in the approximation of
smallness of $\hat{\alpha}$. \\

\textit{Deflection of light rays in presence of gravity and plasma, exact expression.}
In the book of Perlick \cite{Perlick2000} the method was developed for consideration of a deflection of the light rays in presence of the gravity and plasma. The general formulae for the exact light deflection angle in the Schwarzschild and Kerr metric, in presence of plasma with spherically symmetric distribution of concentration, are obtained in the form of integrals. The case with a Schwarzschild metric is described in the next section. Formula (\ref{main-res}) can be also obtained from the exact formula, in a weak deflection limit.

In the present paper we give another derivation of the formula for an exact photon deflection angle by a Schwarzschild metric, in presence of a plasma with a
spherically symmetric distribution of concentration, using Synge's approach. We present a convenient
way to express the integral for gravitational deflection angle in the case of a homogeneous plasma via elliptic integrals. We derive also the asymptotic analytic formula for the gravitational deflection angle in the homogeneous plasma, in a strong deflection limit. We apply this formula for calculation of angular positions and magnification factors of relativistic images in homogeneous plasma.\\

\textit{Deflection of light rays in different media in presence of gravitation.}
Let us summarize the main properties of the gravitational deflection in different media.

(i) In a vacuum the gravitational deflection is achromatic.

(ii) If the medium is homogeneous, the refractive index $n$ is constant in space, and the  medium is not dispersive, so that a refractive index does not depend on the photon frequency, then the
gravitational deflection angle is the same as in the vacuum. A light group velocity in this case is smaller than the light velocity in vacuum.

(iii) If the medium is homogeneous but dispersive, so that the refractive index is constant in  space but depends on the wave frequency, the gravitational deflection angle is different from
the vacuum case and depends on the photon frequency. The example of this case is a homogeneous plasma.

(iv) If the medium is non-homogeneous, so that the refractive index depends on the space coordinates, we will have also the refractive deflection. If medium is non-homogeneous and dispersive, the refractive deflection depends on the photon frequency.
\\

\textit{Analogy between a photon in a plasma and a massive particle in a vacuum.}
Plasma has unique dispersive properties. In the paper of Kulsrud and Loeb \cite{Kulsrud-Loeb} it was shown that in the homogeneous
plasma the photon wave packet moves like a particle with a velocity equal to the group velocity of the wave packet $v_{gr} =
\sqrt{1-\frac{\omega_e^2}{\omega^2}}$,  with a mass  $m_{pl} = \hbar \omega_e$, and with an energy
$E_{pl} = \hbar \omega$. We have shown \cite{BKTs2010} that our result for the light deflection in the weak deflection limit (\ref{main-res}) can be
obtained using this analogy. Due to this analogy, some of results obtained in the present paper can be applied to the motion and deflection of massive
particles in vacuum.

\section{Exact deflection angle in plasma}

In this section we derive the exact expression for the deflection angle in the case of a photon motion in the Schwarzschild
metric with a spherically symmetric distribution of plasma. We use spherical coordinates $(r, \theta, \varphi)$ and the gravitational
field is not supposed to be weak. Indexes are

\begin{equation}
i,k = 0,1,2,3; \quad \alpha, \beta = 1,2,3 \; (r, \theta, \varphi) \, ,
\end{equation}
\begin{equation}
\mbox{ signature } \; \; \{-,+,+,+\} .
\end{equation}
Let us consider a static space-time with the Schwarzschild metric:
\begin{equation}
ds^2 = g_{ik} dx^i dx^k = - A(r) \, dt^2 + \frac{dr^2}{A(r)} + r^2
\left( d \theta^2 + \sin^2 \theta d \varphi^2 \right),
\label{metric}
\end{equation}

\begin{equation}
\mbox{where }  \;\; A(r) = 1 - \frac{2M}{r} \, .
\end{equation}
Let us consider, in this gravitational field, a static inhomogeneous plasma with a refraction index (\ref{plasma-n}).

The general relativistic geometrical optics in a curved space-time, in a dispersive medium with an angular isotropy of the refraction index, was developed by Synge
\cite{Synge}. It is based on the so called equation of medium for the static case, giving the
connection between the phase velocity $w$, and a 4-vector of the
photon momentum $p^i$. Using  the refraction index of the medium
$n$, $n=1/w$, this connection is written as
\begin{equation} \label{eq-medium}
n^2 = 1 + \frac{p_i p^i}{\left(p_0 \sqrt{-g^{00}}\right)^2} \, .
\end{equation}
In this equation the metric $g_{ik}$ and the refraction index $n$ are assumed to be known, the refractive index $n$ is a function of
$x^\alpha$ and $\omega(x^\alpha)$. For a plasma $n$ is given in (\ref{plasma-n}). Equation (\ref{eq-medium}) connects the photon energy and 3-vector of the photon momentum, in given medium and in presence of gravitational field. For a static medium in a static gravitational field, we have \cite{Synge}:

\begin{equation} \label{Synge}
p_0 \sqrt{-g^{00}} = - p^0 \sqrt{-g_{00}} = - \hbar \omega(x^\alpha) \, ,
\end{equation}
where $\hbar$ is the Planck constant. In particular, at infinity (in a flat space-time) we have:

\begin{equation} \label{Synge2}
p_0 = - p^0  = - \hbar \omega \, ,
\end{equation}
where $\omega \equiv \omega(\infty)$.
The trajectories of photons, in presence of a gravitational
field, may be obtained from the variational principle \cite{Synge}

\begin{equation} \label{var-princ}
\delta \left(\int p_i \, dx^i\right) = 0 \, ,
\end{equation}
with the restriction  (\ref{eq-medium}), which may be written in
the form

\begin{equation} \label{add-cond}
H(x^i,p_i) = \frac{1}{2} \left[ g^{ik} p_i p_k - (n^2-1) \left(p_0 \sqrt{-g^{00}}\right)^2 \right] = 0 \, .
\end{equation}
Here we define the scalar function $H(x^i,p_i)$ depending on $x^i$ and $p_i$. The variational principle (\ref{var-princ}), with the
restriction $H(x^i,p_i)=0$, leads to the following system of differential equations \cite{Synge}:

\begin{equation}
\label{D-Eq} \frac{dx^i}{d \lambda} = \frac{\partial H}{\partial p_i}  \, , \; \; \frac{dp_i}{d \lambda} = - \frac{\partial
H}{\partial x^i} \, ,
\end{equation}
with the parameter $\lambda$ changing along the light trajectory.
In the case of a plasma with the refractive index
(\ref{plasma-n}), the restriction (\ref{add-cond}) can be reduced,
with using of (\ref{plasma-n}) and (\ref{Synge}), to the form

\begin{equation}
H(x^i,p_i) = \frac{1}{2} \left[ g^{ik} p_i p_k + \omega_e^2 \hbar^2 \right] = 0 \, . \label{H-definition}
\end{equation}
From (\ref{D-Eq}) we obtain the system of equations for the space
components $x^\alpha$, $p_\alpha$:

\begin{equation} \label{eq-motion-x-general}
\frac{dx^\alpha}{d \lambda} = g^{\alpha \beta} p_\beta \, ,
\end{equation}
\begin{equation} \label{eq-motion-p-general}
\frac{dp_\alpha}{d \lambda} = -\frac{1}{2} \, g^{ik}_{,\alpha} p_i p_k - \frac{1}{2} \, \hbar^2 \left( \omega_e^2 \right)_{,
\alpha} \, .
\end{equation}
From the equation in the static metric (\ref{metric}) for the time component $\frac{dp_0}{d \lambda} = 0$, it follows that $p_0 = $ const\, along the trajectory. The equation
(\ref{Synge}) at infinity has the form (\ref{Synge2}), so we find that the constant $p_0$ equals to $p_0= - \hbar \omega$.
Let us find the equation of the trajectory, and the photon deflection angle, for a motion in the equatorial plane $\theta = \pi /2$ of the metric (\ref{metric}).

For a motion in the plane $\theta=0$ we have $p_\theta = p^\theta = 0$. From the equation  (\ref{eq-motion-p-general}) for $p_\varphi$ it
follows  $p_\varphi = \mbox{const}$, see also \cite{MTW}. Without a loss of generality we can assume that $p_\varphi>0$.
The equations for $r$ and $\varphi$ from (\ref{eq-motion-x-general}) may be written as

\begin{equation}
\frac{dr}{d\lambda} = g^{rr} p_r \, , \quad  \frac{d\varphi}{d\lambda} = g^{\varphi \varphi} p_\varphi \, .
\end{equation}
Substituting the components of metric we obtain:

\begin{equation} \label{dphidr-prelim}
\frac{d \varphi}{dr} = \frac{p_\varphi}{r^2 } \, \frac{1}{p_r A(r)} \, .
\end{equation}
The second multiplier in the right hand side of this equation can be expressed from the medium equation for plasma
(\ref{H-definition}), which may be written in the form

\begin{equation}
g^{rr} p_r^2  + g^{\varphi \varphi} p_\varphi^2 + g^{00} p_0^2 + \hbar^2 \omega_e^2(r) =
\end{equation}
\[=A(r) p_r^2  + \frac{p_\varphi^2}{r^2}  - \frac{p_0^2}{A(r)}  + \hbar^2 \omega_e^2(r) = 0 \, .
\]
We have then

\begin{equation} \label{prAr}
p_r A(r) = \pm \sqrt{p_0^2 - A(r)  \left( \frac{p_\varphi^2}{r^2} + \hbar^2 \omega_e^2(r) \right)} \, .
\end{equation}
Substituting (\ref{prAr}) into (\ref{dphidr-prelim}), we obtain the equation of trajectory of photon:

\begin{equation} \label{eq-traekt-general}
\frac{d \varphi}{dr} = \pm \frac{p_\varphi}{r^2 } \, \frac{1}{\sqrt{p_0^2 - A(r)  \left( \frac{p_\varphi^2}{r^2} + \hbar^2
\omega_e^2(r) \right)}} \, .
\end{equation}
Assume that a photon moves in such a way that its $\varphi$-coordinate increases. Then 'plus' sign in (\ref{eq-traekt-general})
corresponds to the motion with the coordinate $r$ also increasing, and the 'minus' sign to the motion with $r$ decreasing.

For a photon which moves from the infinity to the distance of the closest approach $R$ (minimal value of the coordinate $r$), and then to infinity,
we have a change of angular coordinate as

\begin{equation}
\Delta \varphi = - \int \limits_\infty^R \frac{p_\varphi}{r^2 } \, \frac{dr}{\sqrt{p_0^2 - A(r)  \left(\frac{p_\varphi^2}{r^2}
+ \hbar^2 \omega_e^2(r) \right)} }
\end{equation}
$$+ \int \limits_R^\infty \frac{p_\varphi}{r^2 } \, \frac{dr}{\sqrt{p_0^2 - A(r)  \left(
\frac{p_\varphi^2}{r^2} + \hbar^2 \omega_e^2(r) \right)} }
$$$$
=
2 \int \limits_R^\infty \frac{p_\varphi}{r^2 } \, \frac{dr}{\sqrt{p_0^2 - A(r)  \left(\frac{p_\varphi^2}{r^2} + \hbar^2 \omega_e^2(r) \right)} } \, .
$$
The motion along a straight line corresponds to the change of the angular coordinate $\Delta \varphi = \pi$.
Then the deflection angle may be written as

\begin{equation} \label{angle-general}
\hat{\alpha} = 2 \int \limits_R^\infty \frac{p_\varphi}{r^2 } \, \frac{dr}{\sqrt{p_0^2 - A(r)  \left(
\frac{p_\varphi^2}{r^2} + \hbar^2 \omega_e^2(r) \right)} }  - \pi \, .
\end{equation}
The deflection angle in (\ref{angle-general}) depends on a mass of the central body $M$, a
distribution of plasma $N(r)$, represented by $\omega_e(r)$, and on the parameters $R$, $p_0$ and $p_\varphi$. These parameters
are related to the boundary condition. The point $r=R$ is a turning point,
therefore in this point: $dr/d \lambda = 0$ and $p_r = 0$. From (\ref{prAr}) we have in this point

\begin{equation} \label{boundary-general}
p_0^2 = A(R)  \left( \frac{p_\varphi^2}{R^2} + \hbar^2 \omega_e^2(R) \right) \, .
\end{equation}
Therefore only two parameters from \{$R$, $p_0$, $p_\varphi$\} are independent, while the third one is expressed through two others.
The parameter $p_0$ represents the photon energy at infinity. It is convenient to exclude $p_\varphi$, and derive the
angle $\hat{\alpha}$ as a function of $p_0 = - \hbar \omega$ and $R$.  We have from (\ref{boundary-general})

\begin{equation}
p_\varphi^2 = R^2 p_0^2 \left( \frac{1}{A(R)} - \frac{\omega_e^2(R)}{\omega^2}  \right) \, .
\end{equation}
Substituting $p_\varphi^2$ in (\ref{angle-general}) and using notation of Perlick \cite{Perlick2000}

\begin{equation} \label{perlick-h}
h(r) = r \sqrt{ \frac{1}{A(r)} - \frac{\omega_e^2(r)}{\omega^2} } = r \sqrt{ \frac{r}{r-2M} - \frac{\omega_e^2(r)}{\omega^2} } \, ,
\end{equation}
we obtain the equation of trajectory in the Schwarzschild space-time

\begin{equation} \label{perlick-trajectory}
\frac{d \varphi}{dr} = \pm \frac{1}{\sqrt{r(r-2M)}\sqrt{\frac{h^2(r)}{h^2(R)} - 1 }} \, ,
\end{equation}
and the deflection angle of the photon moving from infinity to central object and then to infinity

\begin{equation} \label{perlick-angle}
\hat{\alpha} = 2 \int \limits_R^\infty \frac{dr}{\sqrt{r(r-2M)}\sqrt{\frac{h^2(r)}{h^2(R)} - 1 }} - \pi \, .
\end{equation}
This expression for $\hat\alpha$ was derived earlier in \cite{Perlick2000} by an another way. Under given $M$ and $\omega_e(r)$, the deflection angle is determined by the closest approach distance $R$ and the photon
frequency at infinity $\omega$. This formula allows us to calculate the deflection angle of the photon moving in the Schwarzschild
metric in presence of a spherically symmetric distribution of plasma.

\section{Homogeneous plasma}

Let us consider a homogeneous plasma with $\omega_e(r)=\omega_e =$ const.
Rewriting the equation of the trajectory (\ref{eq-traekt-general}) and the expression for deflection angle (\ref{angle-general}), introducing
notations of $E$ and $L$

\begin{equation} \label{definition-E-L}
\frac{-p_0}{\hbar \omega_e} = \frac{\omega}{\omega_e} = E > 1 \, , \quad \frac{p_\varphi}{\hbar \omega_e} = L > 0 \, ,
\end{equation}
we obtain the equation of the trajectory as

\begin{equation} \label{trajectory-EL}
\frac{d \varphi}{dr} = \pm \frac{L}{r^2} \frac{1}{\sqrt{E^2 - A(r) \left( 1 + \frac{L^2}{r^2} \right) }} \, ,
\end{equation}
and the formula for deflection angle as

\begin{equation} \label{angle-EL}
\hat{\alpha} =  2 \int \limits_R^\infty \frac{L}{r^2} \frac{dr}{\sqrt{E^2 - A(r) \left( 1 + \frac{L^2}{r^2}  \right) }} - \pi \, .
\end{equation}
The closest approach distance $R$ and the parameters $E$ and $L$ are connected at the point $r=R$ by the following boundary condition

\begin{equation} \label{boundary-cond-EL}
E^2 = A(R) \left( 1 + \frac{L^2}{R^2} \right) \, .
\end{equation}
The trajectory and the deflection angle are thus completely determined by any two parameters from \{$R$,  $E$, $L$\}, with the
third parameter being expressed through (\ref{boundary-cond-EL}). Remind, that in vacuum the photon motion
is determined by only one parameter, either $R$, or $b$, which are uniquely connected with each other.
If we define $E$ as an energy at infinity per unit rest mass, and $L$ as the angular momentum per unit rest mass, then equations
(\ref{trajectory-EL}) and (\ref{angle-EL}) describe the trajectory and the deflection angle of a massive particle with the rest mass $\hbar \omega_e$ in the vacuum, see \cite{Kulsrud-Loeb, MTW, Weinberg, Zeld-Novikov}).

\section{Expression of the deflection angle via elliptic integrals}

Here we express the integral for the angle (\ref{angle-EL}) via elliptic integrals. As we mentioned, certain properties of a photon in plasma are
analogous to the corresponding properties of a massive particle in vacuum, therefore, an expression for the deflection angle of a
massive particle in vacuum is identical to (\ref{angle-EL}), with an appropriate changes in the physical sense of $E$ and $L$. The motion of a massive
particle has been extensively studied, see \cite{Hagihara1931,Darwin1959,Darwin1961,Bogorodsky1962, M-Pleb1962,Metzner1963,
Zeld-Novikov,MTW,Chandra,MiroRodriguez1,MiroRodriguez2}. In particular, the deflection angle of a
massive particle can be expressed via elliptic integrals. Substituting $u=1/r$, we find the deflection angle as an integral of
the polynomial of the third order in a radicand, see below. The form of the expression via elliptic integrals is
determined by the roots of this polynomial. In a general case, these three roots are different,
and supposed to be found numerically, for given the external parameters $E$ and $L$. Depending on $E$ and $L$, the roots have
different relations between each other, and the expression of the deflection angle via elliptic integrals have therefore different
form. In the newtonian limit these forms are expressed analytically, and are  related to hyperbolic, parabolic, and elliptic motion of a massive particle around a gravitating center (Kepler problem), see \cite{LLmech}. Here we cannot write the deflection angle as an explicit
function of $E$ and $L$. For given $E$ and $L$, one has to obtain numerically the roots of the polynomial before being able to
compute the deflection angle.

We suggest to use, as independent, another pair of parameters: $R$ and $E$. Then one root of the polynomial is simply $1/R$, and
another two roots can be expressed analytically via $R$ and $E$. This approach is convenient because we can perform all
calculations analytically and obtain the deflection angle in the form of elliptic integrals as an explicit function of two
parameters $R$ and $E$. The proposed novel technique can be also applied for a massive particle in vacuum, due to its analogy to
a photon in plasma. A similar approach is usually applied for the deflection of photons in vacuum, where an analogous polynomial
is expressed via $R$.

The problem is that $R$ is not an {\it ad hoc} value and thus should not be chosen as an external parameter in applications.
This difficulty can be avoided in strong deflection limit, where it becomes possible to express  the deflection angle in terms of
$b$ and $E$ what is convenient for applications. In general case we may use the relation (\ref{boundary-cond-EL}) for expressing $R(E,L)$, where $(E,L)$ are constants defined at infinity.

Let us rewrite (\ref{boundary-cond-EL}) as
\begin{equation} \label{bond-cond-L}
L^2 = R^2 \left( \frac{E^2}{A(R)} - 1 \right) \, ,
\end{equation}
and substitute (\ref{bond-cond-L}) into the expression for the deflection angle (\ref{angle-EL}), to get deflection angle as a function
of $R$ and $E$. Let
us also introduce notations

\begin{equation}
\frac{1}{r} = u, \; \frac{1}{R} = u_0, \; A(u) = 1 - 2Mu, \; A(u_0) = 1 - 2Mu_0,
\end{equation}
and rewrite (\ref{angle-EL}) in terms of $u$ instead of $r$. We will refer the expression in radicand in (\ref{angle-EL}) as
$f(u)$. We obtain (\ref{angle-EL}) in the form:

\begin{equation}
\label{alpha}
\hat{\alpha} =  2 \int \limits_0^{u_0} \frac{L du}{\sqrt {f(u)}} - \pi \, ,
\end{equation}
where

\begin{equation}
f(u) = E^2 - A(u) \left[ 1+ R^2u^2 \left( \frac{E^2}{A(u_0)} - 1 \right) \right].
\end{equation}
Let us write $f(u)$ in the form

\begin{equation}
f(u) = \frac{f_1(u)}{R-2M} \, ,
\end{equation}
where

\begin{equation}
f_1(u) = E^2 (R-2M) -
\end{equation}
$$
- (1-2Mu) (R-2M+R^3 u^2 E^2 - R^3 u^2 + 2MR^2 u^2) \, .
$$
The expression $f_1(u)$ is a polynomial of the third order

$$
f_1(u) = 2M R^2 (R E^2 - R+2M) \, u^3 \, - \, R^2 (R E^2 - R+2M) \, u^2 \, +
$$
\begin{equation}
+ \, 2M(R-2M) \, u  \, + \, (E^2-1)(R-2M) \, .
\end{equation}
If coefficients of this polynomial would be functions of $E$ and $L$, the roots of this polynomial could be found numerically, or by a very cumbersome analytical solution of the third order algebraic equation \cite{BS}. In our
case the coefficients are functions of $R$ and $E$, and one root $u=\frac{1}{R}$ is evidently visible, if we write $f_1(u)$ as

\begin{equation}
f_1(u) = \left( u - \frac{1}{R} \right) \left[  2M R^2 (RE^2 -R+2M) \, u^2 \, - \right.
\end{equation}
$$
\left. - \, R(R-2M)(RE^2-R+2M) \, u \, - \, R(R-2M)(E^2-1) \right] \, .
$$
Other two roots as functions of $R$ and $E$ can be found easily as a solution of a quadratic equation.
We get finally

\begin{equation}
f(u) = \frac{2M R^2  \, (R E^2 - R+2M)}{R-2M}  \, (u-u_A) (u-u_B) (u-u_C) \, ,
\end{equation}
where
\begin{equation} \label{u-ABC}
u_A = \frac{R-2M + Q  }{4MR} \, , \; u_B = u_0 = \frac{1}{R} \, , \; u_C = \frac{R-2M - Q  }{4MR} \, ,
\end{equation}

\begin{equation} \label{definition-Q}
Q^2 = (R-2M)^2 + 8M (R-2M) \, \frac{1}{1+ \frac{2M}{R(E^2-1)}} \, .
\end{equation}
Note, that photons propagate in plasma only when their frequency $\omega^2$ exceed $\omega_e^2$. This condition corresponds to the relation $E > 1$, what for a massive particle corresponds to the quasi-hyperbolic motion (pure hyperbolic in the newtonian approximation). For the deflection angle $\hat{\alpha}$ we have the expression

\begin{equation}
\hat{\alpha} = \frac{2}{\sqrt{2M}} \int \limits_0^{u_0} \frac{du}{\sqrt{(u-u_A)(u-u_B)(u-u_C)}}  -  \pi \, .
\label{eq54}
\end{equation}
Using \cite{Gr-Ryzhik}, we write

\begin{equation} \label{Gr-R}
\int \limits_0^{u_B} \frac{du}{\sqrt{(u-u_A)(u-u_B)(u-u_C)}} = \frac{2}{\sqrt{u_A-u_C}} \, F(z, k) \, ,
\end{equation}
$$
z = \sqrt{\frac{(u_A-u_C)u_B}{(u_B-u_C)u_A}} \, , \quad k = \sqrt{\frac{u_B-u_C}{u_A-u_C}} \, ,
$$
$$
[u_A > u_B > 0 \geq u_C] \, .
$$
Here $F(z,k)$ is a elliptic integral of the first kind \cite{Korn, Gr-Ryzhik}:

\begin{equation}
F(z,k) = \int \limits_0^z \frac{dx}{\sqrt{(1-x^2)(1-k^2 x^2)} } =  \int \limits_0^\varphi \frac{d \theta}{\sqrt{1 - k^2 \sin^2 \theta}} =
\end{equation}
$$
=  \tilde{F} (\varphi, k) \, ,  \; F(\sin \varphi, k) \equiv \tilde{F} (\varphi, k) \, ,  \;  x = \sin \theta \, , \;  z=\sin \varphi \, .
$$
Using (\ref{Gr-R}) in (\ref{eq54}) and substituting (\ref{u-ABC}), we obtain the deflection angle $\hat{\alpha}$ in the form

\begin{equation} \label{alpha-exact1}
\hat{\alpha} = 4 \sqrt{\frac{R}{Q}} \, F(y, k)  -  \pi  \, ,
\end{equation}
\begin{equation}
\mbox{where} \quad y = \sqrt{  \frac{8MQ}{( 6M-R+Q )( R-2M+Q )} }  \, ,
\label{z}
\end{equation}
$$
k = \sqrt{ \frac{6M-R+Q}{2Q} } \, .
$$
This expression can be also written in another form, using of property of elliptic integrals \cite{Abramowitz}, in our
notations:

\begin{equation}
F(z,k) + F(y,k) = F(1,k) \, ,
\label{zy}
\end{equation}
$$
\mbox{provided} \quad \sqrt{1-k^2} \, \frac{z}{\sqrt{1-z^2}} \, \frac{y}{\sqrt{1-y^2}} = 1 \, .
$$
Here $F(1,k) = \tilde{F}(\pi/2,k) = K(k)$ is the complete elliptic integral of the first kind. It is easy to check, that

\begin{equation}
\label{z2}
z^2 = \frac{2M+Q-R}{6M+Q-R} \, .
\end{equation}
and $y$ from (\ref{z}) satisfy the relation  (\ref{zy}). Therefore,
we obtain for the deflection angle $\hat{\alpha}$ in the form

\begin{equation} \label{alpha-exact2}
\hat{\alpha} = 4 \sqrt{\frac{R}{Q}} \, \left[ F(1,k) - F(z,k) \right]   -  \pi.
\end{equation}
Therefore we have obtained, for the first time, the formulae for the deflection angle (\ref{alpha-exact1}), (\ref{alpha-exact2}), for massive bodies, where arguments of elliptical integrals are expressed explicitly via parameters $R$ and $E$, defining the trajectory.

This formula is written in the same form as the expression for the vacuum
deflection angle, see \cite{MTW}, \cite{Chandra}, \cite{IyerPetters}, \cite{BKTs2008}. The difference between plasma and vacuum
formulae is only in the expression for $Q$. When $E$ goes to infinity, what corresponds to high energy photons, for which plasma effects are negligible, the expression for $Q$ in (\ref{definition-Q}) goes to  $Q^2 =
(R-2M)(R+6M)$, and the formula (\ref{alpha-exact2})  transforms into the formula for the vacuum deflection.
The expression (\ref{alpha-exact1}) for the vacuum case with $Q^2 = (R-2M)(R+6M)$ is written as
\cite{Darwin1959}

\begin{equation}
\hat{\alpha} = 4 \sqrt{\frac{R}{Q}} \, F(y, k)  -  \pi   \,  ,
\end{equation}
\begin{equation}
\mbox{where} \quad y = \sqrt{  \frac{2Q}{3R-6M+Q} }  \, , \quad k = \sqrt{ \frac{6M-R+Q}{2Q} }  \, .
\end{equation}

\section{Trapped photon trajectories in plasma around a black hole}
As for massive particle in vacuum, bound elliptic orbits of the photon in homogeneous plasma are also possible. Gravitational binding of the photon in homogeneous plasma was also discussed in the paper of Kulsrud and Loeb \cite{Kulsrud-Loeb}.

For the existence of a bound orbit it is necessary to have a  photon frequency higher than the plasma frequency, at all points of the trajectory. Otherwise, the photon will be absorbed by plasma. We need:
\begin{equation}
\omega(r) \ge \omega_e \, , \quad \mbox{or} \quad E(r) = \frac{\omega(r)}{\omega_e} \ge 1 \, .
\end{equation}
Here $\omega(r)$ is locally measured frequency. We remind that in notations of this paper $\omega$ and $E$ are constants: $\omega \equiv \omega(\infty)$, $E \equiv E(\infty) $. We have (compare with \cite{MTW}):
\begin{equation}
\omega(r) = \sqrt{-g^{00}} \, \omega \, , \quad E(r) = \sqrt{-g^{00}} \, E \, .
\end{equation}
\begin{equation}
E^2(r) = \frac{E^2}{A(r)} \, .
\end{equation}

At all points of trajectory we have (see (\ref{trajectory-EL}) and (\ref{angle-EL})):
\begin{equation}
E^2 \ge A(r) \left( 1 + \frac{L^2}{r^2} \right) \, .
\end{equation}
We obtain:
\begin{equation}
E^2(r) = \frac{E^2}{A(r)} \ge  1 + \frac{L^2}{r^2} \ge 1 \, .
\end{equation}

If $E<1$, then at some $r=R_1$ there will be
$$
E^2(R_1) = 1 + \frac{L^2}{R_1^2} \, ,
$$
and the photon cannot penetrate outside $r=R_1$. Only trapped photons, inside the radius $R_1$, may survive in these conditions.

\section{Critical distance of the closest approach for unbound photons}

To obtain an  expression in the limit of a strong deflection angle we need to expand the integral (\ref{alpha-exact2}) around the
value of $R$, at which the deflection angle goes to infinity. This value is a function of $E = \omega/\omega_e$, $\omega \equiv \omega(\infty)$, and we'll call it as the critical value of $R$.
The deflection angle (\ref{angle-EL}) can be written in terms of an effective potential similarly to the motion of
massive particles in vacuum \cite{MTW}
\begin{equation}
\hat{\alpha} = 2 \int \limits_0^{u_0} \frac{L du}{\sqrt{E^2 - V^2(u)}} - \pi \, ,
\end{equation}
where
\begin{equation}
V^2(u) = (1-2Mu) ( 1 + L^2 u^2 ) \, .
\end{equation}
Here $L$ is defined in (\ref{definition-E-L}). From

\begin{equation}
\frac{dV^2(u)}{du} = 0
\end{equation}
we obtain

\begin{equation}
u = \frac{1 \pm \sqrt{1-12M^2/L^2}}{6M}  \, .
\end{equation}
The maximum and minimum of the effective potential are situated at the points $r_M$ and $r_m$ respectively:

\begin{equation} \label{rM_L}
r_M = \frac{6M}{1 + \sqrt{1-12M^2/L^2}} \, ,
\end{equation}
$$
r_m = \frac{6M}{1 - \sqrt{1-12M^2/L^2}} \, , \quad r_M < r_m \, .
$$
Point $R=r_M$ corresponds to unstable circular orbit of the photon, $r_m$ corresponds to stable circular orbit (see also Fig.\ref{rm-rM}). The critical distance for the closest approach is reached when the value of $R$ coincide with $r_M$. Using (\ref{bond-cond-L}) and (\ref{rM_L}) we express the critical distance $r_M$ for unbounded photons coming from infinity, as a function of $E$ in the form

\begin{equation} \label{rM_E}
r_M = \frac{M}{2}\left[\left(3-\frac{1}{u_1}\right)+\frac{1}{u_1}\sqrt{1+10u_1+9u_1^2}\right],\,\,
\end{equation}
\[L^2=\frac{M\,r_M^2}{r_M-3M},\,\,
u_1 \equiv E^2-1>0.\]
For a "parabolic" motion with $E=1,\,\, u_1=0$, corresponding to infinitely long wave photons with infinitely small energy at infinity, we have $r_M=4M$, $L^2=16M^2$. For very energetic photons  $E\rightarrow \infty$ we have $r_M\rightarrow 3M$, like for the photons in the vacuum. In this case (\ref{rM_E})
$L\rightarrow \infty$ also. That corresponds to ultrarelativistic limit of the massive particle. Therefore we have that for unbounded photons with  $E>1$
it should be $L^2 > 16 M^2$ to avoid absorption by a black hole, as for massive particle in vacuum \cite{MTW}.
When $R$ tends to $r_M$ ($R>r_M$), the deflection angle goes to infinity. To obtain the analytic expression for the strong
deflection angle we need to expand the integral (\ref{alpha-exact2}) around $R=r_M$. Expressing $r_M$ via $E$ in (\ref{rM_E}), we obtain

\begin{equation}
r_M = \frac{3E^2 - 4 + 3E^2 x}{2(E^2-1)} \, M \, , \; x \equiv \sqrt{1-\frac{8}{9E^2}} = \sqrt{1-\frac{8 \omega_e^2}{9 \omega^2}} \, .
\end{equation}
Noticing that $3E^2-4+3E^2x =\frac{3}{2} E^2 (3x-1)(1+x)$, we write $r_M$ as

\begin{equation}
r_M = \frac{3}{4} \, \frac{(3x-1)(1+x)}{1-\frac{1}{E^2}} \, M \, .
\end{equation}
And using the equality $E^2(3x-1)(3x+1)=8(E^2-1)$, we have for $r_M$

\begin{equation} \label{rM-x}
r_M = 6M \, \frac{1+x}{1+3x} \, .
\end{equation}
Under given ratio of frequencies $\omega_e^2/\omega^2$, the trajectory is determined by a choice of $R$. The deflection angle of a photon goes to infinity when $R$ goes to $r_M$, and photon performs infinite number of turns at the radius $r_{M}$. Therefore $r_{M}$ is a critical (minimal) distance of the closest approach $R$, for a given $\omega_e^2/\omega^2$.
If $\omega \gg \omega_e$, then $x \to 1$, giving $r_{M} = 3M$, what corresponds to the photons in the vacuum.
For unbound orbits of massive particles coming from infinity, making several loops around a black hole,
and going back to infinity, the value of a critical distance of the closest approach also lays between $3M$ and $4M$.
So, the presence of a homogeneous plasma increases the radius of the critical orbits of the photons around a black hole.

The conception of a photon sphere has been introduced by Virbhadra and Ellis \cite{Virbhadra2000} as a timelike hypersurface of the form ${r=r_0}$ where $r_0$ is the closest distance of approach for which the Einstein bending angle of a light ray is unboundedly large. These authors subsequently \cite{Virbhadra2002} considered the Einstein deflection angle for a general static spherically symmetric metric and obtained the equation defining the photon sphere. In any space–time containing the photon sphere, gravitational lensing will give rise to relativistic images \cite{Virbhadra2000, Virbhadra2001, Virbhadra2002}. In our notations $r_M$ is the distance of closest approach, which is the same as a photon sphere of  \cite{Virbhadra2000, Virbhadra2002}. Therefore the presence of the homogeneous plasma increases the radius of the photon sphere, and this radius depends on the photon frequency.

\section{Deflection angle in a strong deflection limit}

Let us obtain an analytic expansion for the deflection angle $\hat{\alpha}$ in the strong deflection limit, using (\ref{alpha-exact2}).
It follows from (\ref{definition-Q}), (\ref{z}), that $k \simeq 1$ corresponds to $R \simeq r_M$. So we can use expansions of elliptic integrals near $k=1$
\cite{Zhuravsky1941, Gr-Ryzhik}:

\begin{equation}
F(1,k) \simeq \ln \frac{4}{\sqrt{1-k^2}} \, ,
\end{equation}

\begin{equation}
F(z,k) \simeq \ln \tan \left( \frac{\arcsin(z)}{2} + \frac{\pi}{4} \right) \, .
\end{equation}
The expression (\ref{alpha-exact2}) for the deflection angle is simplified to

\begin{equation} \label{alpha-prelim}
\hat{\alpha} = 4 \sqrt{\frac{R}{Q}} \left[ \ln \frac{4}{\sqrt{1-k^2}} - \ln \frac{1+z}{\sqrt{1-z^2}}  \right] - \pi \, .
\end{equation}
Expanding ($1-k^2$) near $R=r_M$ we have, after significant simplifications (see Appendix A)
\begin{equation} \label{1-k2}
1-k^2 \simeq 2 \, \frac{1+x}{x} \, \frac{M(R-r_M)}{r_M^2} \, .
\end{equation}
Thus, expanding (\ref{alpha-prelim}) near $R=r_M$ and keeping leading terms, we obtain:
\begin{equation} \label{alpha-with-z}
\hat{\alpha} = -2 \sqrt{\frac{1+x}{2x}} \ln \left[ \frac{1+x}{8x} \, \frac{(R-r_M)M}{r_M^2} \, \frac{(1+z)^2}{1-z^2}  \right] -
\pi \, ,
\end{equation}
where $z$ from (\ref{z2}) is simplified to
\begin{equation}
z = \sqrt{\frac{3x-1}{6x}} \, .
\end{equation}
Substituting $z$, we obtain the deflection angle $\hat{\alpha}$ as
a function of the closest approach distance $R$ and the ratio of frequencies
$\omega_e^2/\omega^2$ in the form

\begin{equation} \label{alpha-R-E}
\hat{\alpha}(R,x) = -2 \sqrt{\frac{1+x}{2x}} \ln \left[z_1(x) \, \frac{R-r_M}{r_M}  \right] - \pi \, ,
\end{equation}
where
\begin{equation} \label{z-x}
\quad z_1(x) = \frac{9x-1+2\sqrt{6x(3x-1)}}{48x} \, ,
\end{equation}

\begin{equation} \label{r-M}
r_M=6M \, \frac{1+x}{1+3x} \, , \; x \equiv \sqrt{1-\frac{8}{9E^2}} = \sqrt{1-\frac{8 \omega_e^2}{9 \omega^2}} \, .
\end{equation}
This formula (\ref{alpha-R-E}) is asymptotic and valid for $R$ close to $r_M$. Under $\omega \gg \omega_e$ the expression (\ref{alpha-R-E}) becomes the vacuum formula (\ref{vacuum-alpha-R}).
The impact parameter $b$ for a massive particle is defined uniquely by the energy at infinity, per unit rest mass of particle, $E$ and the
angular momentum, per unit rest mass of particle, $L$ as \cite{MTW}, \cite{Weinberg}:

\begin{equation} \label{definition-b}
b^2 = \frac{L^2}{E^2-1} \, .
\end{equation}
So in case of motion of photon in plasma we will have the same expression for impact parameter, if we use our expressions for $E$
and $L$ from (\ref{definition-E-L}). Since we need $b$ as a function of $R$ and $E$, we substitute $L^2$ from (\ref{bond-cond-L})
into (\ref{definition-b}). Let $b_{cr}$ is the minimal impact parameter, corresponding to $r_M$. Expanding $b$ in (\ref{definition-b}) around $R=r_M$, we obtain:
\begin{equation}
b = b_{cr} + b_1 (R-r_M)^2 \, ,
\end{equation}
where
\begin{equation} \label{b-cr}
b_{cr} = \sqrt{3} \, r_M \, \sqrt{\frac{1+x}{3x-1}} \, ,
\end{equation}
\begin{equation}
b_1 = \frac{3\sqrt{3}}{2} \, \frac{x}{r_M} \, \sqrt{\frac{1+x}{3x-1}} = \frac{3x}{2r_M^2} \, b_{cr} \, .
\end{equation}
Writing:
\begin{equation}
\frac{(R-r_M)^2}{r_M^2} = \frac{2}{3x} \, \frac{b-b_{cr}}{b_{cr}} \, ,
\end{equation}
we obtain the deflection angle $\hat{\alpha}$ as a function of the impact parameter $b$ and the ratio of frequencies
$\omega_e^2/\omega^2$:
\begin{equation} \label{alpha-b-E}
\hat{\alpha}(b,x) = - \sqrt{\frac{1+x}{2x}} \, \ln \left[ \frac{2\,z_1^2(x)}{3x} \, \frac{b-b_{cr}}{b_{cr}} \right] - \pi \, .
\end{equation}
This formula is valid for $b$ close to $b_{cr}$, where $b_{cr}$ is a critical value of impact parameter under given $\omega_e^2/\omega^2$. For $\omega \gg \omega_e$ we obtain the critical
impact parameter for vacuum: $b_{cr} = 3 \sqrt{3} M$. Under $\omega \gg \omega_e$ the expression (\ref{alpha-b-E}) becomes the vacuum
formula (\ref{vacuum-alpha-b}).

\section{Positions of relativistic images}

Formula (\ref{alpha-b-E}) can be applied for calculation of positions of the relativistic images in presence of homogeneous
plasma (see Figure 4). For simplicity, let us rewrite formula (\ref{alpha-b-E}) as:
\begin{equation} \label{alpha-simple}
\hat{\alpha}(b,x) = - a(x) \, \ln \left( \frac{b-b_{cr}}{b_{cr}} \right) + c(x) \, ,
\end{equation}
where $a(x)$ and $c(x)$ are coefficients
\[
a(x) = \sqrt{\frac{1+x}{2x}} \, , \quad c(x) = - \sqrt{\frac{1+x}{2x}} \, \ln \left[ \frac{2\,z_1^2(x)}{3x} \right] - \pi \, ,
\]
$z_1(x)$ is defined in (\ref{z-x}), $b_{cr}$ is defined in (\ref{b-cr}) with $r_M$ and $x$ from (\ref{r-M}).

To obtain the impact parameters corresponding to the relativistic images we can write an equation:
\begin{equation} \label{2-pi-n}
\hat{\alpha}(b,x) = 2 \pi n \, , \quad n=1,2, ...
\end{equation}
Here $n$ is number of pair of relativistic images. To be rigorous, the deflection angle in this case is $\hat{\alpha} = 2 \pi +
\Delta \hat{\alpha}$ (see Figures 2 and 4), where $\Delta \hat{\alpha}$ has different values for different positions of source. But since
$\Delta \hat{\alpha} \ll 2 \pi n$ we can write equation for $b$ in the form (\ref{2-pi-n}), with high accuracy. We obtain the
impact parameters $b_n(x)$ of the relativistic images:
\begin{equation}
b_n = b_{cr} \left[ 1 + \exp \left(\frac{c(x)-2\pi n}{a(x)} \right) \right] \,  .
\end{equation}
The corresponding angular positions of the relativistic images:
\begin{equation} \label{theta-n}
\theta_n = \frac{b_{cr}}{D_d} \left[ 1 + \exp \left(\frac{c(x)-2\pi n}{a(x)} \right) \right]  \,  ,
\end{equation}
where $D_d$ is the distance between observer and lens.

The angular positions $\theta_n$ in plasma is always bigger than the positions in vacuum. In Table 1 results for $E = 2$ ($\omega =2 \omega_e$) are presented. Therefore presence of homogeneous plasma \textit{increases} the angular separation of the point relativistic images from gravitating center or the angular size of the relativistic rings.

\begin{table}[b]
\caption{\label{table1}
Comparison of the impact parameters and the angular sizes of the relativistic rings in vacuum and homogeneous plasma for $\omega = 2 \omega_e$ ($E =2$).  For convenience, differences $b_n-b_{cr}$ are presented. The values of impact parameters are given in units of mass $M$. The angular sizes can be calculated as $\theta = b/D_d$. These values can also be considered as positions of two relativistic images of the point source, in case if there is no perfect alignment.}
\begin{ruledtabular}
\begin{tabular}{lcc}
Relativistic rings & vacuum & plasma \\
\colrule
Critical value ($b_{cr}$) & $3\sqrt{3}$ & $3\sqrt{3} + 0.540300928156$ \\
Ring 1 ($b_1-b_{cr}$) & 0.00650 & $0.00899$ \\
Ring 2 ($b_2-b_{cr}$) & $1.21 \cdot 10^{-5}$ & $2.05 \cdot 10^{-5}$ \\
Ring 3 ($b_3-b_{cr}$) & $2.27 \cdot 10^{-8}$ & $4.68 \cdot 10^{-8}$ \\
\end{tabular}
\end{ruledtabular}
\end{table}

Let us consider $b_n$ near the vacuum values. With $E^2 \rightarrow \infty$ ($\omega \gg  \omega_e$) we have:
\begin{equation}
r_M \simeq 3M \left( 1 + \frac{1}{9E^2}  \right) , \; b_{cr} \simeq 3 \sqrt{3} M \left(  1 + \frac{1}{3E^2} \right) ,
\end{equation}
\begin{equation}
\theta_n = \frac{b_n}{D_d} \, , \quad   b_n \simeq 3\sqrt{3} M (1+l_n^{vac})  \,  +
\end{equation}
$$
+ \, 3\sqrt{3} M \frac{1}{9E^2} \left\{ 3 + l_n^{vac} \left[ \pi(2n+1) + 2\sqrt{3} -3   \right] \right\} \, ,
$$
where
\begin{equation} \label{e-n-v}
l_n^{vac} = 216 (7 - 4\sqrt{3}) \, e^{-\pi (2n+1)} \, .
\end{equation}
Note that $E=\infty$ correspond to vacuum.

Let us consider a behavior of $b_n$ in opposite situation: when the photon frequency is close to plasma frequency. With $E^2 \rightarrow 1$ ($\omega \simeq  \omega_e$) we have:
\begin{equation}
r_M \simeq 4M [ 1 + (E^2-1)  ]   \, ,
\end{equation}
\begin{equation}
b_{cr} \simeq \frac{4M}{\sqrt{E^2-1}} \simeq \frac{4M}{n_{pl}}  \; \left( \mbox{here} \; \; n_{pl}=\sqrt{1-\frac{\omega_e^2}{\omega^2}} \right) ,
\end{equation}
\begin{equation}
\theta_n = \frac{b_n}{D_d} \, , \; b_n \simeq \frac{4M}{\sqrt{E^2-1}}  \left[  1+  32 \, e^{-\pi (2n+1)/ \sqrt{2}} \right] .
\end{equation}
We see that with $\omega \rightarrow  \omega_e$ angular sizes $\theta_n$ can increase unboundedly.

\section{Magnifications of relativistic images}

Here we consider the magnification of the relativistic images of a point source located at the angular position $\beta$ from line connecting the observer and the gravitational center (lens). We use approximate approach, similar to \cite{Bozza2002}, based on our analytical formula (\ref{alpha-simple}). More precise way is to use exact formula (\ref{perlick-angle}) and solve this problem by the similar approach as it was done in \cite{Virbhadra2000} for relativistic images in vacuum.


To derive the magnification factors of the relativistic images in our approach, we need to write a lens equation and take into account that the deflection angle $\hat{\alpha}$ of rays forming the relativistic images is slightly different from $2\pi n$. We write $\hat{\alpha}(b)$ near $b=b_n$ as
\begin{equation}
\hat{\alpha}(b) \simeq \hat{\alpha}(b_n) + \left. \frac{\partial \hat{\alpha}}{\partial b} \right|_{b=b_n} (b-b_n) = 2 \pi n +  \Delta \hat{\alpha}_n \, .
\end{equation}
Using formula (\ref{alpha-simple}), we obtain
\begin{equation}
\Delta \hat{\alpha}_n = - a(x) \frac{b-b_n}{b_n-b_0} \, .
\end{equation}
Now we can write the lens equation as (see \cite{Bozza2001}, \cite{Bozza2002}):
\begin{equation}
\beta = \theta - \frac{D_{ds}}{D_s} \Delta \hat{\alpha}_n \, ,
\end{equation}
where $\beta$ is the angular position of source, $\theta$ is the angular position of image, $\Delta \hat{\alpha}_n$ is a function of $\theta$ (or $b$), $n$ is number of pair of relativistic images, see Fig.2. This form of the lens equation is approximate, for more exact lens equation see \cite{Virbhadra2000}, \cite{Frittelli2000}, \cite{Perlick2004a}. We obtain:
\begin{equation}
\beta = \theta + \frac{D_{ds}}{D_s} a(x) \frac{\theta-\theta_n}{\theta_n-\theta_0} \, .
\end{equation}
The magnification factors of the relativistic images located at the angular positions $\theta_n$ are defined as
\begin{equation}
\mu_n = \left. \left( \frac{\beta}{\theta}  \frac{\partial \beta}{\partial \theta} \right)^{-1} \right|_{\theta=\theta_n} = \frac{\theta_n}{\beta} \left. \left(  \frac{\partial \beta}{\partial \theta} \right)^{-1} \right|_{\theta=\theta_n} \, .
\end{equation}
We obtain:
\begin{equation} \label{mu-n}
\mu_n = \frac{D_s b_{cr}^2 (1+ l_n) l_n}{D_{ds} D_d^2  \, a(x) \, \beta} \, ,
\end{equation}
where
\begin{equation}
l_n = \exp \left(\frac{c(x)-2\pi n}{a(x)} \right) ,
\end{equation}
and $a(x)$ and $c(x)$ are coefficients defined in (\ref{alpha-simple}). In expression (\ref{mu-n}) the variables $b_{cr}$, $l_n$, $a(x)$, $c(x)$ depend on $x$, so these variables depend on the ratio of the photon and the plasma frequencies. Magnification $\mu_n$ of the relativistic images tends to infinity if the source angular position $\beta$ goes to 0, as it takes place in vacuum \cite{Virbhadra2000}.

In case of vacuum ($\omega/\omega_e=\infty$):
\begin{equation} \label{mu-n-v}
\mu_n^{vac} = \frac{D_s b_{cr}^2 (1+ l_n^{vac}) l_n^{vac}}{D_{ds} D_d^2  \, \beta} \, , \quad b_{cr} = 3\sqrt{3} M \, ,
\end{equation}
where $l_n^{vac}$ is defined in (\ref{e-n-v}).

If $E^2 \rightarrow 1$  ($\omega \simeq \omega_e$), we have:
\begin{equation}
\mu_n = \frac{D_s}{D_{ds} D_d^2} \, \frac{16M^2}{E^2-1}   \,  \frac{(1+l_n^{pl})  l_n^{pl}}{\sqrt{2} \, \beta} \, ,
\end{equation}
where
\begin{equation}
l_n^{pl} = 32 e^{-\pi(2n+1)/\sqrt{2}} \, .
\end{equation}
We see that if $\omega \rightarrow \omega_e$, magnifications $\mu_n$ can increase unboundedly.

Numerical examples are given in Table 2.

The magnification of relativistic rings may be calculated in presence of plasma by the same method, as it was done for vacuum in \cite{BKTs2008}.

\begin{table}[b]
\caption{\label{table1}
Comparison of the magnification factors of relativistic images for lensing in homogeneous plasma and in vacuum. Ratios $\mu_n/\mu_n^{vac}$ are presented, at given $D_s$, $D_d$, $D_{ds}$, $\beta$. Values $\mu_n$ and $\mu_n^{vac}$ are calculated with (\ref{mu-n}) and (\ref{mu-n-v}) correspondingly. Values $\mu_n$ and $\mu_n^{vac}$ depend on $D_s$, $D_d$, $D_{ds}$, $\beta$, $\omega/\omega_e$, but ratio $\mu_n/\mu_n^{vac}$ depends only on $\omega/\omega_e$. In last column the values $\mu_n^{vac}$ are calculated for $M/D_d = 2.26467 \cdot 10^{-11}$, what corresponds to supermassive black hole in center of Milky Way, $D_s/D_{ds}=2$, $\beta = 1 \mu$as (these parameters have been taken from \cite{Virbhadra2009}).    }
\begin{ruledtabular}
\begin{tabular}{lcccc}
$\mu_n / \mu_n^{vac}$ &
$\omega/\omega_e=1.1$ &
$\omega/\omega_e=2$   &
$\omega/\omega_e=10$  &
$\mu_n^{vac}$ \\
\colrule
$\mu_1 / \mu_1^{vac}$  &  $13.3$ & 1.48 & 1.01 & $ 0.716 \cdot 10^{-11}$ \\
$\mu_2 / \mu_2^{vac}$  &  $40.0$ & 1.81 & 1.02 & $ 0.134 \cdot 10^{-13}$ \\
$\mu_3 / \mu_3^{vac}$  &  $120$  & 2.21 & 1.03 & $ 0.249 \cdot 10^{-16}$ \\
\end{tabular}
\end{ruledtabular}
\end{table}

\section{Duscussion}

We can make the following conclusions:

(i) Using Synge's approach for the geometrical optic in the curved space-time, in dispersive medium \cite{Synge}, we find the expression for the deflection angle of the photon moving in the Schwarzschild metric in plasma with the spherically symmetric density distribution, obtained in \cite{Perlick2000} by another way, see (\ref{perlick-angle}). For the homogeneous plasma we express the deflection angle via elliptic integrals, see (\ref{alpha-exact1}) and (\ref{alpha-exact2}).

(ii) We derive formula for the deflection angle of the photon moving in the Schwarzschild metric in the homogeneous plasma, in strong deflection limit. Formula is derived as a function of $(R,\omega_e/\omega)$ and $(b, \omega_e/\omega)$, see (\ref{alpha-R-E}) and (\ref{alpha-b-E}) correspondingly.

(iii) We calculate the angular positions and the magnifications of the relativistic images, see (\ref{theta-n}) and (\ref{mu-n}). We conclude that in presence of the homogeneous plasma the angular positions and the magnifications of the relativistic images increase, in comparison with the vacuum case. If the photon frequency $\omega$ is close to the plasma frequency $\omega_e$, the angular positions and the magnifications formally increase significantly, but it corresponds to very long radio waves, where absorption is large.\\

These results can be used for investigation of relativistic images in the case of lensing by supermassive black holes in the center of galaxies. There are many difficulties to observe such relativistic images, but present results can be considered as a big step to possible future radio observation of relativistic images. Possibility and difficulties of observation of relativistic images are discussed in big details in \cite{Virbhadra2000} and \cite{Virbhadra2009}.

Our formulae may be also applied in investigations of the X-ray binary containing a black hole. The light from the companion may have a strong deflection angle for sufficiently small impact parameters.\\

In this work we pay our attention mainly to the case of a homogeneous plasma. This case allows us to get good analytical results.
In reality the plasma in the neighborhood of the compact objects can be significantly non-homogeneous. Such cases can be
calculated numerically with using of the formula (\ref{perlick-angle}). Another important thing is that plasma near the compact
object is rather moving than static. Taking into account the motion of the medium requires the constructing of a more general
model, which should also allow us to describe the Doppler effect.

\section*{Acknowledgments}

We would like to thank an anonymous referee for motivation to more deep investigation of the paper subject.

The work of GSBK and OYuT was partially supported by the Russian Foundation for Basic Research grant 11-02-00602, the RAN Program 'Formation and evolution of stars and galaxies', and the Russian Federation President Grant for Support of Leading Scientific Schools NSh-5440.2012.2.

The work of OYuT was also partially supported by the Russian Foundation for Basic Research grant 12-02-31413 (project 'My first grant'), the Russian Federation President Grant for Support of Young Scientists
MK-2918.2013.2 and the Dynasty Foundation.

\appendix

\section{Some calculations for Section VIII}
To find an expansion (\ref{1-k2}) of $1-k^2$ near $R=r_M$, let us rewrite $k^2$ as
\begin{equation} \label{zA-1}
k^2 = \frac{6M-R+Q}{2Q} = \frac{1}{2} + \frac{1}{2} \, \frac{6M-R}{Q} \, .
\end{equation}
For convenience, let us calculate $Q^2/(6M-R)^2$. Here $Q^2$ is
\begin{equation} \label{zA-def-Q}
Q^2 = (R-2M)^2 + 8M (R-2M) \, \frac{1}{1+ \frac{2M}{R(E^2-1)}} \, .
\end{equation}
Let us write $R$ as
\begin{equation}
R = r_M + \Delta R \, .
\end{equation}
Here $\Delta R = R - r_M$ is a small variable:
\begin{equation}
\Delta R \ll r_M \, ,
\end{equation}
and
\begin{equation} \label{zA-rM-x}
r_M = 6M \, \frac{1+x}{1+3x} \, .
\end{equation}

Expanding $Q^2$ near $R=r_M$ (small variable is $\Delta R$), we obtain:
\begin{equation}
Q^2 \simeq Q_0 + Q_1 \, \Delta R \, ,
\end{equation}
where we define the notations
\begin{equation} \label{zA-def-Q0}
Q_0 = (r_M - 2M)^2 + 8M \, \frac{r_M - 2M}{z_a} \, ,
\end{equation}
\begin{equation} \label{zA-def-Q1}
Q_1 = 2 (r_M-2M) + \frac{8M}{z_a} +  \frac{(r_M-2M) \, 16M^2}{z_a^2 (E^2-1) \, r_M^2} \, ,
\end{equation}
\begin{equation} \label{zA-def-za}
z_a = 1 + \frac{2M}{(E^2-1) \, r_M} \, .
\end{equation}

Expanding $(6M-R)^2$ near $R=r_M$ (small variable is $\Delta R$), we obtain:
\begin{equation}
(6M-R)^2 \simeq (r_M - 6M)^2 \left( 1 + \frac{2}{r_M-6M} \Delta R \right) \, .
\end{equation}

Keeping leadings terms, we obtain $Q^2/(6M-R)^2$ in the form
\begin{equation}
\frac{Q^2}{(6M-R)^2} \simeq f_0 + f_1 \, \Delta R \, ,
\end{equation}
where we define the notations
\begin{equation} \label{zA-def-f0}
f_0 = \frac{Q_0}{(r_M-6M)^2} \, ,
\end{equation}
\begin{equation} \label{zA-def-f1}
f_1 = \frac{1}{(r_M-6M)^2} \left( Q_1 - \frac{2Q_0}{r_M-6M}  \right) \, .
\end{equation}
We need to simplify $f_0$ and $f_1$.

Let us simplify $z_a$. Substituting $r_m$ into $z_a$ we obtain:
\begin{equation}
z_a = 1 + \frac{2M}{E^2-1} \frac{1}{6M} \frac{1+3x}{1+x}
\end{equation}
Using
\begin{equation} \label{zA-3x}
E^2 (3x+1) (3x-1) = 8(E^2-1)
\end{equation}
we obtain:
\begin{equation}
z_a = 1 + \frac{8}{3E^2} \frac{1}{(3x-1)(1+x)} \, .
\end{equation}
Using
\begin{equation}
9E^2(1+x)(1-x) = 8, \mbox{ or } \frac{8}{E^2(1+x)} = 9(1-x)
\end{equation}
we obtain
\begin{equation} \label{zA-za}
z_a = \frac{2}{3x-1} \, .
\end{equation}
With substituting $r_M$ from (\ref{zA-rM-x}), expressions $(r_M-2M)$ and $(r_M-6M)$ can be written as
\begin{equation} \label{zA-rM2M-rM6M}
r_M - 2M = \frac{4M}{1+3x}, \quad r_M - 6M = - \frac{12Mx}{1+3x} \, .
\end{equation}

Using (\ref{zA-za}) and (\ref{zA-rM2M-rM6M}) in (\ref{zA-def-Q0}), we get:
\begin{equation} \label{zA-Q0-rM6M}
Q_0 = \frac{144M^2x^2}{(1+3x)^2} = (r_M-6M)^2 \, .
\end{equation}
Using (\ref{zA-Q0-rM6M}) in (\ref{zA-def-f0}) we obtain that $f_0=1$.

Let us simplify $f_1$ from (\ref{zA-def-f1}). Using (\ref{zA-Q0-rM6M}) we obtain:
\begin{equation}
f_1 = \frac{1}{(r_M-6M)^2} \left( Q_1 - 2(r_M-6M)  \right) \, .
\end{equation}
Substituting $Q_1$, we obtain:
\begin{equation}
f_1 = \frac{8M}{(r_M-6M)^2} \left( 1 + \frac{1}{z_a} + \frac{(r_M-2M) \, 2M}{z_a^2 (E^2-1) \, r_M^2}  \right) \, .
\end{equation}
Using (\ref{zA-za}) and (\ref{zA-rM2M-rM6M}), after simplifications, we obtain
\begin{equation} \label{zA-f1-f2}
f_1 = 8M \, \frac{(1+3x)^2}{144M^2x^2} \, f_2 \, ,
\end{equation}
where
\begin{equation}
f_2 = \frac{3x+1}{2} \, \frac{9(E^2-1)(1+x)^2 + (3x-1)^2 }{9(E^2-1)(1+x)^2}
\end{equation}
We can simplify
\begin{equation}
9(E^2-1)(1+x)^2 + (3x-1)^2 = 6x (3E^2 + 3E^2x -4) \, .
\end{equation}
Noticing that
\begin{equation}
\frac{3}{2} E^2 (3x-1)(1+x) = 3E^2 + 3E^2x -4 \, ,
\end{equation}
we obtain
\begin{equation}
9(E^2-1)(1+x)^2 + (3x-1)^2 = 9x E^2 (3x-1)(1+x) \, .
\end{equation}
Using (\ref{zA-3x}), after simplifications, we obtain
\begin{equation} \label{zA-f2}
f_2 = \frac{4x}{1+x} \, .
\end{equation}
Substituting $f_2$ to (\ref{zA-f1-f2}) and using (\ref{zA-rM-x}), we obtain
\begin{equation}
f_1 = 8M \, \frac{1+x}{x} \, \frac{1}{r_M^2} \, .
\end{equation}
We have
\begin{equation}
\frac{Q^2}{(6M-R)^2} \simeq 1 + 8M \, \frac{1+x}{x} \, \frac{1}{r_M^2} \, \Delta R \, .
\end{equation}
With using (\ref{zA-1}) we obtain finally
\begin{equation}
1-k^2 \simeq 2 \, \frac{1+x}{x} \, \frac{M}{r_M^2} \, \Delta R \, .
\end{equation}

%

\begin{figure}
\centerline{\hbox{\includegraphics[width=0.45\textwidth]{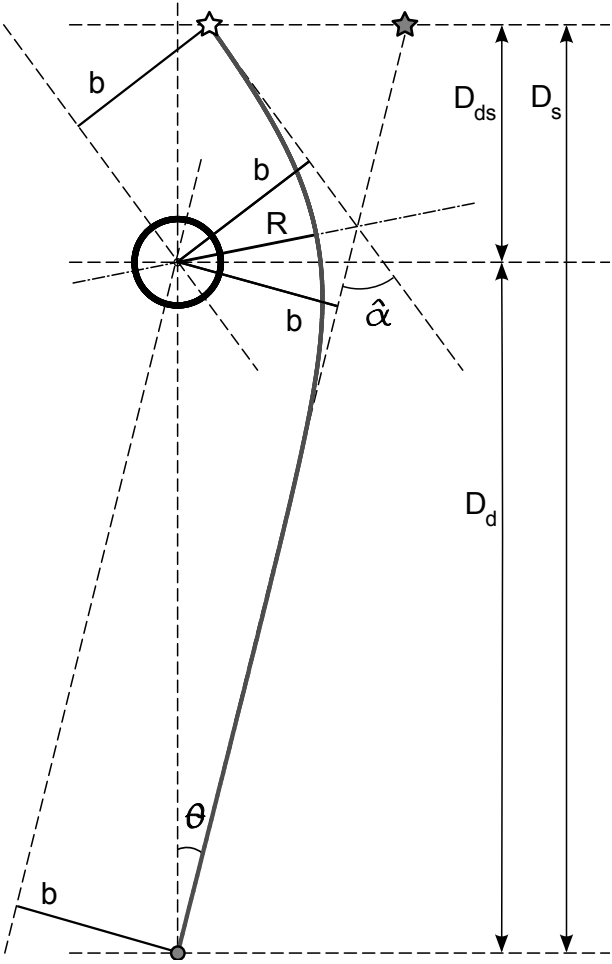}}}  \caption{Gravitational lensing in vacuum. A scheme of
formation of the primary image of the point source. Light ray from the source deflected by angle $\hat{\alpha}$ by the point-mass
gravitational lens with the Schwarzschild radius $R_S$ goes to the observer. The observer sees the image of the source at angular
position $\theta$, which is different from the source position. $R$ is the closest point of trajectory to gravitating center, it
is usually referred as the distance of the closest approach, $b$ is the impact parameter of the photon, $D_d$ is the distance
between observer and lens, $D_s$ is the distance between observer and source, $D_{ds}$ is the distance between lens and source.
The trajectory of the ray is calculated using the equations of the photon motion in vacuum (see, for example, \cite{MTW,
BKTs2008}) for the following values of the parameters: $M=1$, $R_S = 2$, $R = 6.544$, $b \simeq 7.853$, $D_s = 43$, $D_d = 32$,
$D_{ds}=11$.} \label{fig1}
\end{figure}

\begin{figure}
\centerline{\hbox{\includegraphics[width=0.40\textwidth]{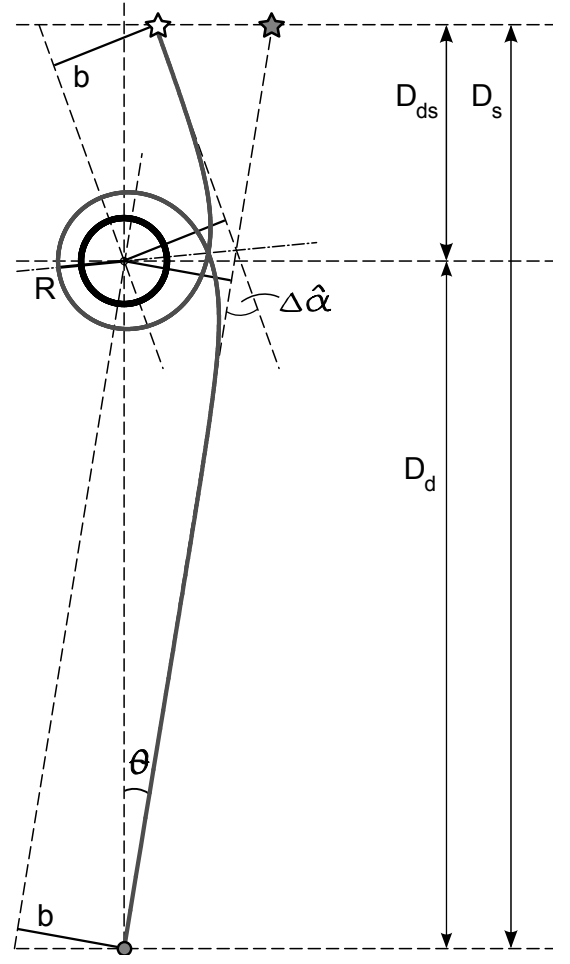}}} \caption{Gravitational lensing in vacuum. A scheme of
formation of the relativistic image of the point source. Light ray from the source is deflected by angle $\hat{\alpha} = 2 \pi +
\Delta \hat{\alpha}$. The impact parameter $b$ is close to its critical value $b_{cr}$. The trajectory of the ray was calculated
using the equations of the photon motion in vacuum (see, for example, \cite{MTW, BKTs2008}) for the following values of the
parameters: $R_S = 2$, $R = 3.068983$, $b \simeq 5.200036$, $D_s = 43$, $D_d = 32$, $D_{ds}=11$.} \label{fig2}
\end{figure}

\begin{figure}
\centerline{\hbox{\includegraphics[width=0.35\textwidth]{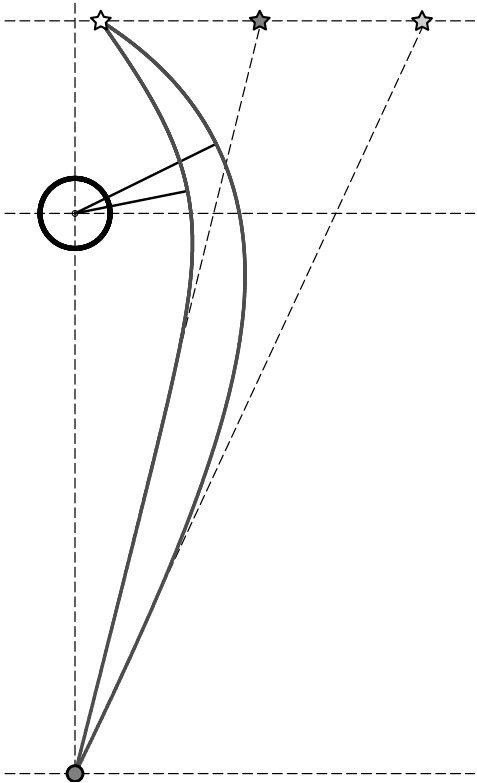}}} \caption{Gravitational lensing in homogeneous plasma. Scheme
of formation of the primary images of different frequencies. The photons of smaller frequency are deflected by a larger angle by
the gravitating center. Instead of the image with all frequencies the observer see a line image, which consists of the images of
different frequencies. At this picture we present only two of such images. The first ray has the frequency $\omega \gg \omega_e$,
in this case plasma effects are negligible, and the trajectory can be computed with using the vacuum equations (see Figure 1,
geometry and positions of objects are the same), $R=6.544$, $r_M=3$. The another ray has the frequency $\omega_e^2/\omega^2 =
9/10$, in this case plasma effects are significant, and the trajectory should be computed with using plasma equations (see
formula (\ref{perlick-trajectory})), $R=8.982$, $r_M = 6(1+x)/(1+3x) \simeq 3.7082$.} \label{fig3}
\end{figure}

\begin{figure}
\centerline{\hbox{\includegraphics[width=0.35\textwidth]{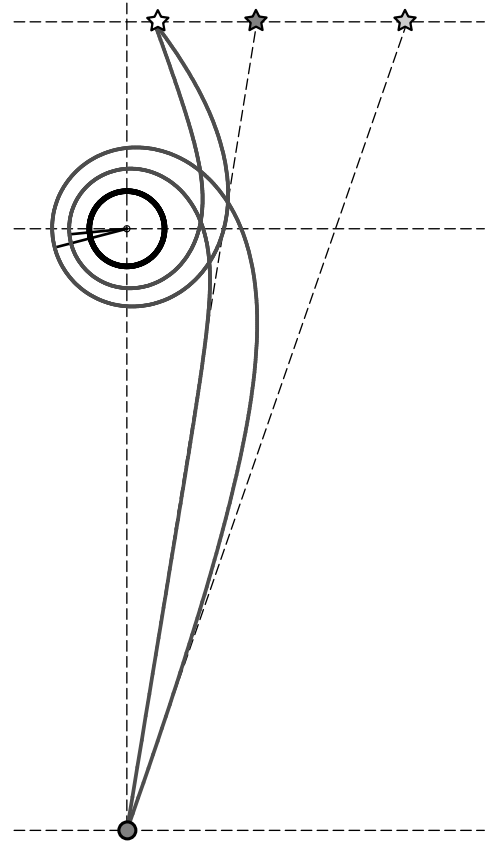}}} \caption{Gravitational lensing in homogeneous plasma. Scheme
of formation of the relativistic images of different frequencies. The photons of smaller frequency are deflected by a larger
angle by the gravitating center. The first ray has the frequency $\omega \gg \omega_e$, in this case plasma effects are
negligible, and the trajectory can be computed with using the vacuum equations (see Figure 1, geometry and positions of objects
are the same), $R=3.068983$, $r_M=3$. The another ray has the frequency $\omega_e^2/\omega^2 = 9/10$, in this case plasma effects
are significant, and the trajectory should be computed with using plasma equations (see formula (\ref{perlick-trajectory})),
$R=3.9675$, $r_M = 6(1+x)/(1+3x) \simeq 3.7082$.} \label{fig4}
\end{figure}

\begin{figure}
\centerline{\hbox{\includegraphics[width=0.48\textwidth]{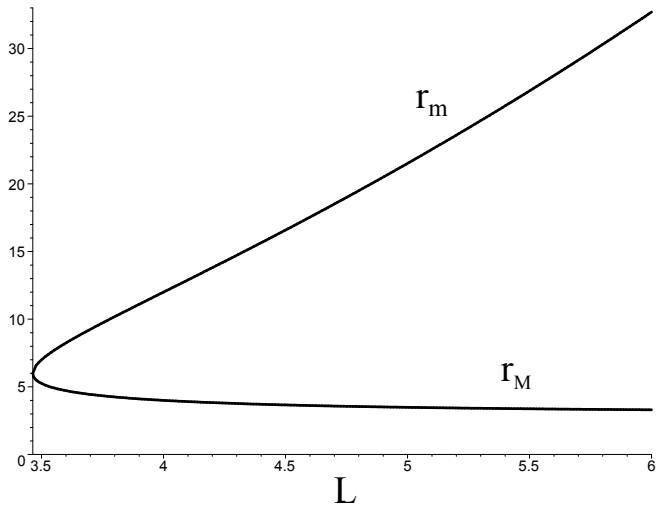}}} \caption{Dependence of $r_m$ and $r_M$ on $L$ (all values are in units of $M$).} \label{rm-rM}
\end{figure}

\end{document}